 \documentclass[smallabstract,smallcaptions]{dccpaper}

\usepackage{amsmath,graphicx,epstopdf}
\usepackage{epsfig}
\usepackage{citesort}
\usepackage{multirow,bigstrut}
\usepackage{amssymb}
\usepackage{color}
\usepackage{url}
\usepackage{booktabs}

\newlength{\figurewidth}
\newlength{\smallfigurewidth}

\begin{document}

\title
{\large
\textbf{A Group Variational Transformation Neural Network for Fractional Interpolation of Video Coding}
}

\author{%
Sifeng Xia, Wenhan Yang, Yueyu Hu, Siwei Ma and Jiaying Liu$^{\ast}$\\[0.5em]
{\small\begin{minipage}{\linewidth}\begin{center}
\begin{tabular}{ccc}
Peking University, Beijing, 100871, China \\
\url{{xsfatpku,yangwenhan,huyy,swma,liujiaying}@pku.edu.cn}
\end{tabular}
\end{center}\end{minipage}}
\thanks{\footnotesize{$^{\ast}$Corresponding author \newline
This work was supported by National Natural Science Foundation of China under contract No. 61772043. We also gratefully acknowledge the support of NVIDIA Corporation with the GPU for this research.}}
}

\maketitle
\thispagestyle{empty}

\begin{abstract}
Motion compensation is an important technology in video coding to remove the temporal redundancy between coded video frames. In motion compensation, fractional interpolation is used to obtain more reference blocks at sub-pixel level. Existing video coding standards commonly use fixed interpolation filters for fractional interpolation, which are not efficient enough to handle diverse video signals well. In this paper, we design a group variational transformation convolutional neural network (GVTCNN) to improve the fractional interpolation performance of the luma component in motion compensation. GVTCNN infers samples at different sub-pixel positions from the input integer-position sample. It first extracts a shared feature map from the integer-position sample to infer various sub-pixel position samples. Then a group variational transformation technique is used to transform a group of copied shared feature maps to samples at different sub-pixel positions. Experimental results have identified the interpolation efficiency of our GVTCNN. Compared with the interpolation method of High Efficiency Video Coding, our method achieves $1.9\%$ bit saving on average and up to $5.6\%$ bit saving under low-delay P configuration.
\end{abstract}
\vspace{-2mm}
\section{Introduction}
Motion compensation is a significant technology in video coding for the temporal redundancy removal between video frames. Specifically, during inter-prediction, there is at least one reference block to be searched from the previously coded frames for each block to be coded. With the reference block, only the motion vector that indicates the position of the reference block and the residual between the blocks need to be coded, which can bring about bit saving in many cases.

However, due to the spatial sampling of digital video, adjacent pixels in a video frame are not continuous, which means that reference blocks at the integer position may not be similar enough to the block to be coded. In order to search for better reference blocks, video coding standards like High Efficiency Video Coding (HEVC) generate reference samples at sub-pixel positions by performing fractional interpolation over the retrieved integer-position sample.

Interpolation methods adopted by the coding standards usually use fixed interpolation filters. For example, MPEG-4 AVC/H.264 \cite{Overviewavc} uses a 6-tap filter for half-pixel interpolation and a simple average filter for quarter-pixel interpolation for luma component. HEVC uses a DCT-based interpolation filter (DCTIF) \cite{Overviewhevc} for fractional interpolation. It is efficient to adopt simple fixed interpolation filters for motion compensation of video coding in real applications. However, the quality of the interpolation results generated by fixed filters may be limited, since fixed filters can not fit for natural and artificial video signals with various kinds of structures and content.

Recently, many deep learning based methods have been proposed for low-level image processing problems, \textit{e.g.} image interpolation \cite{PVLN}, denoising \cite{bddenoise,zhang2017learning}, super-resolution \cite{SRCNN,VDSR,degree,rtsr} and these methods have shown impressive results. In \cite{PVLN}, Yang \textit{et al.} proposed a variational learning network that effectively exploits the structural similarities for image representation. The deep learning based denoising method \cite{bddenoise} utilizes a deep convolutional neural network (CNN) to infer a noise map from the noisy image for denoising. Dong \textit{et al.} proposed a super-resolution method called SRCNN \cite{SRCNN}, which is the first method that uses CNN for super-resolution and has obtained significant gain over traditional super-resolution methods. In \cite{VDSR}, a deeper CNN network with residual learning is built to further improve the super-resolution performance. Besides, edge information is additionally used to guide the inference of high-resolution images in \cite{degree}. Hu \textit{et al.} \cite{rtsr} proposed a global context aggregation and local queue jumping network for image super-resolution considering both reconstruction quality and time consumption.

%Recently, many deep learning based methods have been proposed for low-level image processing problems and they have shown impressive results, \textit{e.g.} image interpolation \cite{PVLN}, denoising \cite{bddenoise,zhang2017learning}, super-resolution \cite{SRCNN,VDSR}. Specifically, Yang \textit{et al.} proposed a variational learning network that effectively exploits the structural similarities for image representation. The deep learning based denoising method \cite{bddenoise} utilizes a deep convolutional neural network (CNN) to infer a noise map from the noisy image for denoising. The super-resolution method called SRCNN \cite{SRCNN} is the first method that uses CNN for super-resolution and has obtained significant gain over traditional super-resolution methods. In \cite{VDSR}, a deeper CNN network with residual learning is built to further improve the super-resolution performance.

Considering the great performance brought by the deep learning based methods in low-level image processing problems and the high implementation efficiency of deep learning based methods brought by GPU acceleration, it is a new opportunity to utilize deep based interpolation methods in motion compensation for video coding. Yan \textit{et al.} \cite{halfpel} first proposed a CNN-based interpolation filter to replace the half-pixel interpolation part of HEVC. Their method has obtained obvious gain over HEVC, which has also demonstrated the superiority of deep learning based interpolation in video coding. However, in their method, only the half-pixel interpolation is replaced and the quarter-pixel interpolation is still the one of HEVC. Furthermore, they chose to train one model for each half-pixel position, which leads to a high storage cost and sets a barrier to the applications in some extreme conditions.

In this paper, we propose a deep learning based fractional interpolation method to infer reference samples of all sub-pixel positions for motion compensation in HEVC. A group variational transformation convolutional neural network (GVTCNN) is designed to infer samples of various sub-pixel positions with one single network. The network firstly extracts a shared feature map from the integer-position sample. Then a group variational transformation method is used to infer different samples from a group of shared feature maps. Experimental results show the superiority of our GVTCNN in fractional interpolation which further benefits video coding.

The rest of the paper is organized as follows. Sec. \ref{sec2} introduces the proposed group variational transformation convolutional neural network based fractional interpolation. Details of fractional interpolation in HEVC are first introduced and analyzed. Then the architecture of the network is illustrated and the group variational transformation is described. Process of the generation of training data is also presented. Experimental results are shown in Sec. \ref{sec3} and concluding remarks are given in Sec. \ref{sec4}.

\begin{figure}[t]
\centering
\includegraphics[width=100mm]{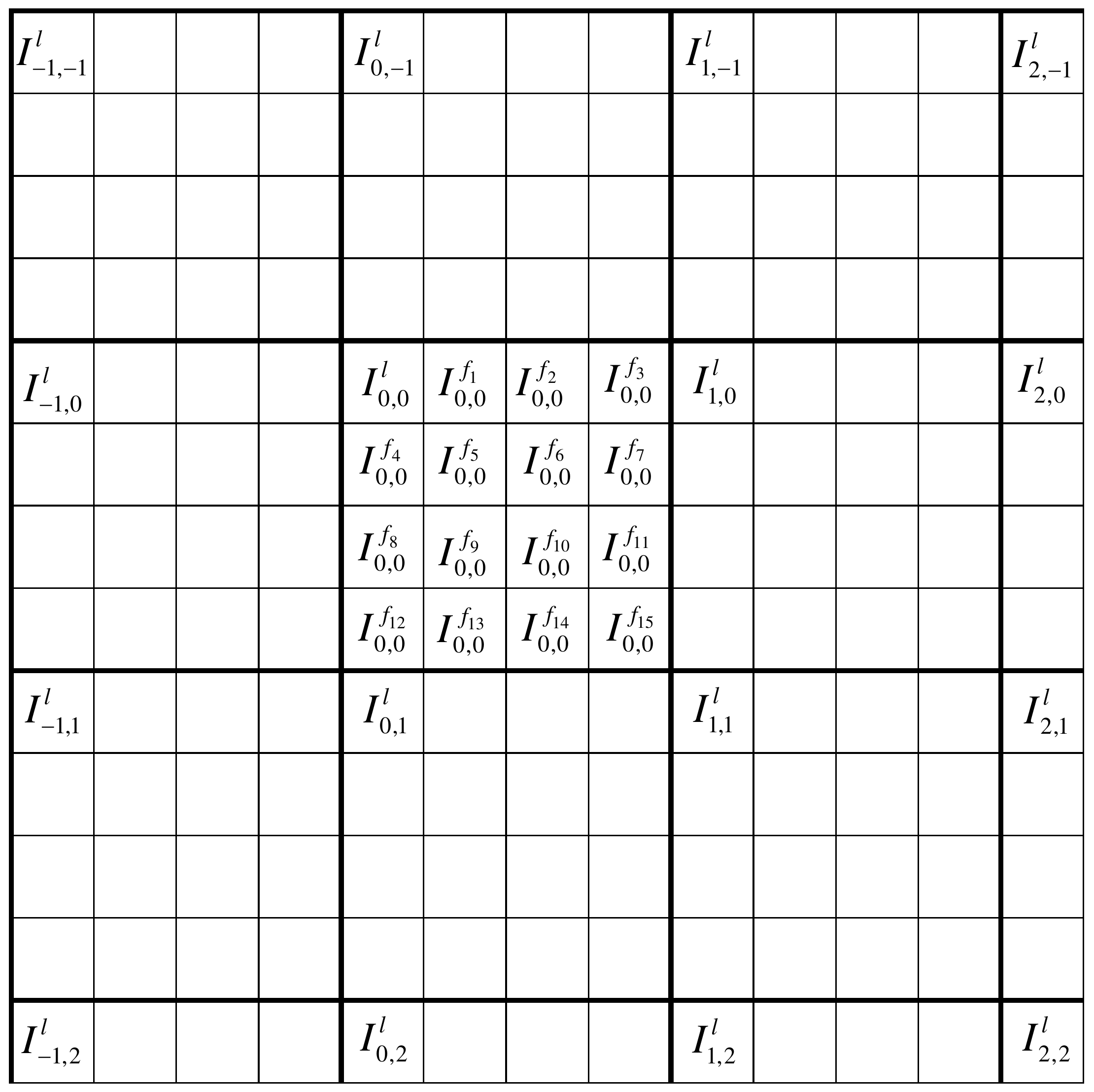}
\caption{Half-pixel and quarter-pixel sample positions in luma component fractional interpolation.}
\label{fig:f1}
\end{figure}

\begin{figure}[t]
\includegraphics[width=160mm]{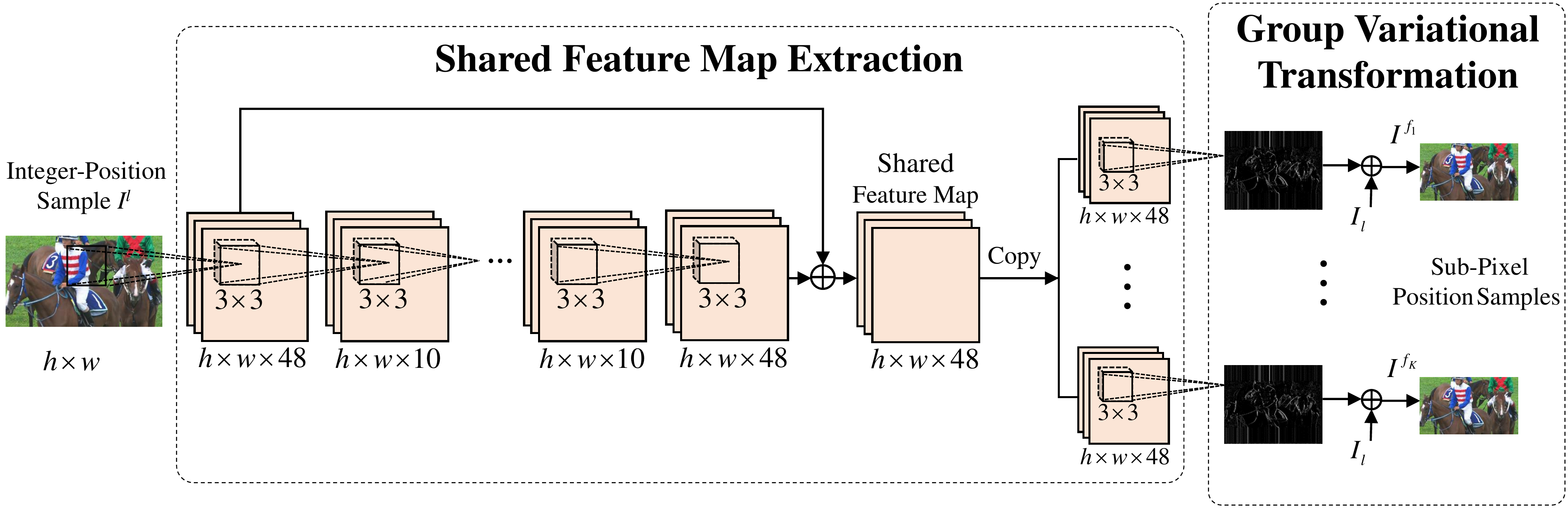}
\caption{Framework of the proposed group variational transformation convolutional neural network (GVTCNN).}
\label{fig:f2}
\end{figure}

\newpage
\section{Group Variational Transformation Neural Network \\ Based Fractional Interpolation}
\label{sec2}

\subsection{Fractional Interpolation in HEVC}

During motion compensation of the luma component in HEVC, at least one integer-position sample in the previously coded frame is first searched for each query block which is to be coded. As shown in Fig. \ref{fig:f1}, half-pixel position samples ${I^{{f_2}}}$, ${I^{{f_8}}}$ and ${I^{{f_{10}}}}$ are first interpolated based on the integer-position sample ${I^{l}}$. And quarter-pixel position samples are then interpolated with the values of integer and half-pixel position samples. The most appropriate reference sample is finally selected among the integer, half-pixel position and quarter-pixel position samples to facilitate coding the query block.

HEVC adopts a uniform 8-tap filter for half-pixel interpolation and 7-tap filters for quarter-pixel interpolation. The fixed interpolation filters may not be flexible enough to accomplish all the interpolation tasks of various kinds of video scenes well. Moreover, the interpolation of each sub-pixel only covers a small area of the integer-position sample, which means that only limited reference information is utilized for sub-pixels generation. As a result, the interpolation results of HEVC may not be good enough in some hard cases like the scenes with complex structures.

Deep convolutional neural network based methods have obtained much success in such kinds of low-level image processing problems. With the help of training data, deep CNN based methods learn a mapping from the input signal $x$ to the target result $y'$ by:
\begin{equation}
\label{eq1}
y' = f\left( {x,\Theta }\right),
\end{equation}
where $\Theta$ represents the set of learnt parameters of the convolutional neural network, which are learnt based on the training data with the back-propagation algorithm.

\subsection{Group Variational Transformation Neural Network}
The proposed GVTCNN consists of two components: the shared feature map extraction part and the group variational transformation part. The shared feature map is first extracted by GVTCNN from the integer-position sample $I^{l}$ and group variational transformation infers the residual maps of the samples at different sub-pixel positions based on the shared feature map.

Fig. \ref{fig:f2} shows the architecture of GVTCNN. The integer-position sample $I^{l}$ is the input of the network. $h \times w \times c$ represents the size of each convolutional layer, where $h$ and $w$ are respectively the height and the width of the feature map, and $c$ is the channel number of the feature map. $3 \times 3$ indicates the size of the convolutional kernal. In this paper, we use $3 \times 3$ kernel size for all the convolutional layers. The parametric rectified linear units (PReLU) \cite{prelu} are utilized for nonlinearity between the convolutional layers.

Specifically, we define $f^{out}_{k}$ to be the output of the $k$-th convolutional layer. $f^{out}_{k}$ is obtained by:
\begin{equation}\label{eq2}
f_k^{out} = {P_k}\left( {{W_k}*f_{k - 1}^{out} + {B_k}} \right),
\end{equation}
where $f_{k - 1}^{out}$ is the output of the previous layer, $W_k$ is the convolutional filter kernel of the $k$-th layer and $B_k$ is the bias of the $k$-th layer. $f_{0}^{out}$ is the input integer-position sample. The function ${P_k}\left( \cdot \right)$ is the PReLU function of the $k$-th layer:
\begin{equation}\label{eq3}
{P_k}\left( x \right) =
\left\{
\begin{aligned}
& x, & &x > 0, \\
& {a_k}*x, & &x \leq 0. \\
\end{aligned}
\right.
\end{equation}
$x$ is the input signal and ${a_k}$ is the parameter to be learned for the k-th layer. ${a_k}$ is initially set as $0.25$ and all channels of the $k$-th layer share the same parameter ${a_k}$.

In the shared feature map extraction component, a feature map with $48$ channels is initially generated from the integer-position sample, followed by $8$ convolution layers with $10$ channels which are lightweight and cost less to save the learnt parameters. The $10$-th layer later derives a $48$ channel feature map. The residual learning technique is utilized in shared feature map extraction for accelerating the convergency of the network. So that we add the $1$-st layer to the $10$-th layer and then activate the sum with PReLU function to obtain the shared feature map. After $9$ convolutional layers with $3 \times 3$ kernel size, the receptive field of each point in the shared feature map is $19 \times 19$, which means that a large nearby area in the integer-pixel position sample has been considered for the feature extraction of each pixel.

Considering the spatial correlation and continuity of the sub-pixels, we argue that there is no need to separately extract a feature map for the generation of each sub-pixel position sample. In other words, we do not need to train a network for each sub-pixel position sample, which is inconvenient for real applications. As a result, after the shared feature map extraction, the shared feature map is used to infer the sub-pixel samples at different locations. The group variational transformation is further performed over the shared feature maps with a specific convolutional layer for each sub-pixel sample. Different residual maps are then generated and we obtain the final inferred sub-pixel position samples by adding the residual maps to the integer-position sample.

During training process, mean square error is used as the loss function. Let $F\left(  \cdot  \right)$ represent the learnt network that infers sub-pixel position samples from the integer-position sample and $\Theta$ denote the set of all the learnt parameters including the convolutional filter kernels, bias and ${a_k}$ of the PReLU function in each layer. The loss function can be formulated as follows:
\begin{equation}\label{eq4}
L\left( \Theta  \right) = \frac{1}{n}\sum\limits_{i = 1}^n {{{\left\| {F\left( {{x_i},\Theta } \right) - {y_i}} \right\|}^2}},
\end{equation}
where pairs $\{x_{i},y_{i}\}_{i=1}^{n}$ are the generated ground-truth pairs of integer-position and sub-pixel position samples and $n$ is the total number of the pairs.

\subsection{Training Data Generation for Motion Compensation}
In the deep CNN based interpolation and super-resolution methods, the ground truth high-resolution images are directly used as the label for the loss function and the down-sampled images are used as the input. However, there are big differences between image resolution recovery problems and fractional interpolation in video coding.

%training data generation of the fractional interpolation in video coding should be further designed.

Firstly, methods of interpolation and super-resolution recover a high-resolution image from the low-resolution image, where ground truth high-resolution images exist. The resolution recovery quality can be measured by simply calculating the differences between the recovered images and the ground truth images. Fractional interpolation in video coding differently aims to generate more sub-pixel position samples for motion compensation. And the efficiency of sub-pixel position samples generation is measured by final coding performance. Secondly, fractional interpolation is performed over the reconstructed previously coded reference frame, therefore the additional information loss brought by the reconstruction in video coding should also be considered in training data generation. Besides, in many video coding standards, half-pixel position interpolation and quarter-pixel position interpolation are performed separately, which is reasonable since these two kinds of interpolation provide reference samples at different sub-pixel levels.

\begin{figure}[t]
\includegraphics[width=140mm]{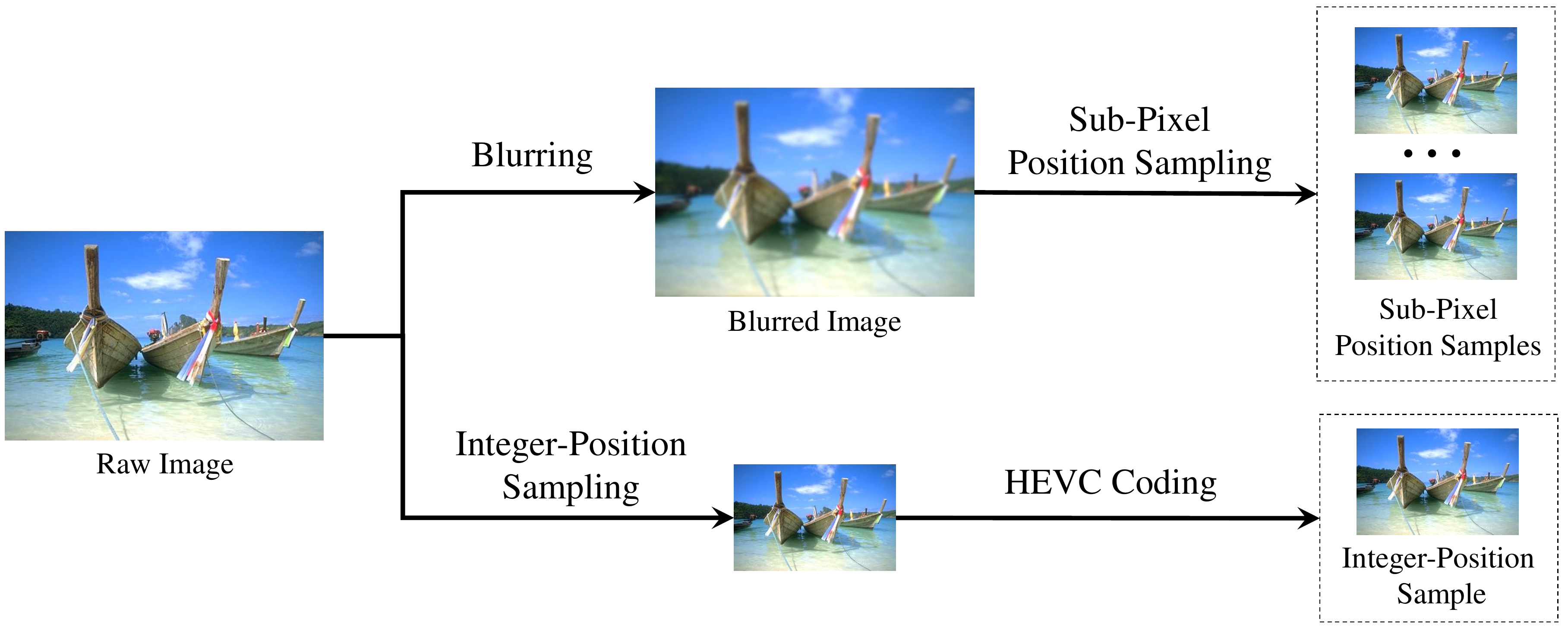}
\caption{Flow chart of the training data generation for GVTCNN.}
\label{fig:f3}
\end{figure}

\begin{table}[!h]
\centering
\footnotesize
\caption{BD-rate reduction of the proposed method compared to HEVC.}
\begin{tabular}{c|l|c|c|c}
\hline
\multirow{2}[4]{*}{Class} & \multirow{2}[4]{*}{Sequence} & \multicolumn{3}{c}{BD-rate} \bigstrut\\
\cline{3-5}      &       & Y     & U     & V \bigstrut\\
\hline
\multirow{6}[12]{*}{Class B} & Kimono & -3.7\% & 0.9\% & 1.3\% \bigstrut\\
\cline{2-5}      & BQTerrace & -5.6\% & -4.2\% & -5.5\% \bigstrut\\
\cline{2-5}      & BasketballDrive & -3.3\% & -0.9\% & -1.1\% \bigstrut\\
\cline{2-5}      & ParkScene & -1.1\% & 0.0\% & -0.7\% \bigstrut\\
\cline{2-5}      & Cactus & -2.2\% & -0.6\% & -0.9\% \bigstrut\\
\cline{2-5}      & Average & -3.2\% & -1.0\% & -1.4\% \bigstrut\\
\hline
\hline
\multirow{5}[10]{*}{Class C} & BasketballDrill & -2.3\% & -1.4\% & -0.7\% \bigstrut\\
\cline{2-5}      & BQMall & -2.8\% & -1.3\% & -1.2\% \bigstrut\\
\cline{2-5}      & PartyScene & -0.6\% & 0.0\% & -0.6\% \bigstrut\\
\cline{2-5}      & RaceHorsesC & -1.9\% & -0.7\% & -1.6\% \bigstrut\\
\cline{2-5}      & Average & -1.9\% & -0.8\% & -1.0\% \bigstrut\\
\hline
\hline
\multirow{5}[10]{*}{Class D} & BasketballPass & -3.1\% & -1.2\% & -1.6\% \bigstrut\\
\cline{2-5}      & BlowingBubbles & -1.6\% & -0.2\% & -0.5\% \bigstrut\\
\cline{2-5}      & BQSquare & 1.6\% & 1.7\% & 2.7\% \bigstrut\\
\cline{2-5}      & RaceHorses & -1.8\% & -1.7\% & -1.6\% \bigstrut\\
\cline{2-5}      & Average & -1.2\% & -0.3\% & -0.2\% \bigstrut\\
\hline
\hline
\multirow{4}[8]{*}{Class E} & FourPeople & -2.2\% & -0.9\% & -0.7\% \bigstrut\\
\cline{2-5}      & Johnny & -2.7\% & -0.4\% & 0.7\% \bigstrut\\
\cline{2-5}      & KristenAndSara & -2.1\% & 0.5\% & 0.4\% \bigstrut\\
\cline{2-5}      & Average & -2.3\% & -0.3\% & 0.1\% \bigstrut\\
\hline
\hline
\multirow{5}[10]{*}{Class F} & BasketballDrillText & -1.7\% & -0.6\% & -0.3\% \bigstrut\\
\cline{2-5}      & ChinaSpeed & -1.4\% & -1.9\% & -1.6\% \bigstrut\\
\cline{2-5}      & SlideEditing & 0.2\% & 0.1\% & 0.0\% \bigstrut\\
\cline{2-5}      & SlideShow & -0.1\% & 0.2\% & -0.6\% \bigstrut\\
\cline{2-5}      & Average & -0.8\% & -0.6\% & -0.6\% \bigstrut\\
\hline
\hline
All Sequences & Overall & -1.9\% & -0.6\% & -0.7\% \bigstrut\\
\hline
\end{tabular}%
\label{tab1}
\end{table}

As a result, we correspondingly take several measures in training data generation to make the trained network applied to fractional interpolation in video coding. The overall flow chart of the process of training data generation has been shown in Fig. \ref{fig:f3}. Referring to the method mentioned in \cite{halfpel}, the training data is first blurred with a gaussian filter to simulate the correlations between the integer-position sample and sub-pixel position samples. Sub-pixel position samples are later sampled from the blurred image. As for the input integer-position sample generation, an intermediate integer sample is previously down-sampled from the raw image. Then, the intermediate down-sampled version is coded by HEVC video coding to simulate the information loss of the reconstructed reference sample.

Moreover, we train two networks separately for $3$ half-pixel position samples and $12$ quarter-pixel position samples to better generate the samples at different sub-pixel levels. And there are some differences between the training data generation of the network that infers half-pixel position samples and the network that infers quarter-pixel position samples (respectively called GVTCNN-H and GVTCNN-Q). In the process of training data generation for GVTCNN-H, $200$ training images and $200$ testing images in the set \emph{BSDS500} \cite{BSD} at size $481 \times 321$ and $321 \times 481$ are used for training. $3 \times 3$ Gaussian kernels with random standard deviations in the range $\left[ {0.5,0.6} \right]$ are used for blurring. By dividing the images into $2\times2$ patches without overlapping, pixels at the top-left of the patches in the raw images are sampled to obtain the integer-position sample. And pixels at other three positions of the patches are separately sampled from the blurred image to derive the sub-pixel position samples.

For GVTCNN-Q, the inferred samples are at a smaller sub-pixel level. The sampling will be performed based on $4\times4$ patches, where $12$ samples are extracted from pixels at $1/4$ or $3/4$ positions vertically or horizontally in the patch. It is not appropriate to continue using images at relatively low resolution for training data generation here, since differences between the integer samples and sub-pixel samples may be a little big with a sampling step length of $4$, which may affect the training of GVTCNN-Q. So images at larger size are chosen to generate the training data to keep the continuity and similarity between the input integer samples and the target sub-pixel position samples. $10$ YUV sequences at size $1024 \times 768$ and $1920 \times 1080$ are used to extract $89$ high resolution frames to synthesize training data\footnotemark[1]. Standard deviations of Gaussian kernels ranges from $\left[ {0.7,0.8} \right]$.

\footnotetext[1]{\url{http://media.xiph.org/video/derf/}}

\section{Experimental Results}
\label{sec3}
\subsection{Experimental Settings}
During the training process, the training images are decomposed into $32\times32$ sub-images with a stride of 16. The GVTCNN is trained on Caffe platform \cite{caffe} via Adam \cite{adam} with standard back-propagation. The learning rate is initially set as a fixed value $0.0001$ and dropped after $30,000$ iterations by a factor of $10$. The batch size is set as $128$. Models after $50,000$ iterations are used for testing. The network is trained on one Titan X GPU.

\begin{table}[!t]
\footnotesize
  \centering
    \caption{BD-rate reduction of CNNIF and the proposed GVTCNN-H for different classes.}
\begin{tabular}{c|l|c|c|c|c|c|c}
\hline
\multirow{2}[4]{*}{Class} &  \multirow{2}[4]{*}{Sequence}     & \multicolumn{3}{c|}{CNNIF\cite{halfpel}} & \multicolumn{3}{c}{GVTCNN-H} \bigstrut\\
\cline{3-8}      &       & Y     & U     & V     & Y     & U     & V \bigstrut\\
\hline
\multirow{5}[10]{*}{Class C} & BasketballDrill & -1.2\% & -0.6\% & 0.2\% & -1.9\% & -1.3\% & -0.4\% \bigstrut\\
\cline{2-8}      & BQMall & -0.9\% & 0.3\% & 0.7\% & -2.0\% & -0.8\% & -0.9\% \bigstrut\\
\cline{2-8}      & PartyScene & 0.2\% & 0.5\% & 0.3\% & -0.3\% & -0.1\% & -0.1\% \bigstrut\\
\cline{2-8}      & RaceHorsesC & -1.5\% & -0.5\% & -0.1\% & -1.6\% & -1.0\% & -0.2\% \bigstrut\\
\cline{2-8}      & Average & -0.9\% & -0.1\% & 0.3\% & -1.4\% & -0.8\% & -0.4\% \bigstrut\\
\hline
\hline
\multirow{5}[10]{*}{Class D} & BasketballPass & -1.3\% & -0.4\% & 0.3\% & -2.4\% & -1.2\% & -0.7\% \bigstrut\\
\cline{2-8}      & BlowingBubbles & -0.3\% & 0.4\% & 0.8\% & -0.9\% & 0.9\% & -0.5\% \bigstrut\\
\cline{2-8}      & BQSquare & 1.2\% & 2.9\% &3.1\% & 1.9\% & 2.0\% & 3.7\% \bigstrut\\
\cline{2-8}      & RaceHorses & -0.8\% & -0.9\% & 0.0\% & -1.1\% & -0.9\% & -0.2\% \bigstrut\\
\cline{2-8}      & Average & -0.3\% & 0.5\% & 1.0\% & -0.6\% & 0.2\% & 0.6\% \bigstrut\\
\hline
\hline
\multirow{4}[8]{*}{Class E} & FourPeople & -1.3\% & -0.4\% & 0.1\% & -2.1\% & -0.5\% & -0.3\% \bigstrut\\
\cline{2-8}      & Johnny & -1.2\% & -0.4\% & -0.7\% & -2.7\% & -1.1\% & -0.6\% \bigstrut\\
\cline{2-8}      & KristenAndSara & -1.0\% & 0.3\% & 0.2\% & -2.3\% & 0.1\% & 0.1\% \bigstrut\\
\cline{2-8}      & Average & -1.2\% & -0.2\% & -0.1\% & -2.4\% & -0.5\% & -0.3\% \bigstrut\\
\hline
\end{tabular}%
  \label{tab2}
\end{table}

The proposed method is tested on HEVC reference software HM 16.15 under the low delay P (LDP) configuration. BD-rate is used to measure the rate-distortion. The quantization parameter (QP) values are
set to be $22$, $27$, $32$ and $37$. We also compare with the CNN based half-pixel interpolation method proposed in \cite{halfpel}. During the training data generation, each intermediate integer sample is coded by four QPs: $22$, $27$, $32$ and $37$. And we train a GVTCNN-H and GVTCNN-Q for each QP based on the corresponding training data. For other QPs, models belong to their nearest QP are chosen. A CU level rate-distortion optimization is also integrated to decide whether to replace our deep based interpolation method with the interpolation method of HEVC.

\begin{table}[!h]
\small
  \centering
    \caption{Average BD-rate reduction achieved by training the networks separately and uniformly.}
\begin{tabular}{c|c|c|c|c|c|c}
\hline
\multirow{2}[4]{*}{Class} & \multicolumn{3}{c|}{GVTCNN-Separate} & \multicolumn{3}{c}{GVTCNN-Uniform} \bigstrut\\
\cline{2-7}      & Y     & U     & V     & Y     & U     & V \bigstrut\\
\hline
Class C & -1.9\% & -1.0\% & -1.2\% & -1.9\% & -0.8\% & -1.0\% \bigstrut\\
\hline
Class D & -1.3\% & -0.1\% & -0.2\% & -1.2\% & -0.3\% & -0.2\% \bigstrut\\
\hline
Class E & -2.4\% & -0.1\% & -0.4\% & -2.3\% & -0.3\% & 0.1\% \bigstrut\\
\hline
\end{tabular}%
  \label{tab3}
\end{table}

\subsection{Experimental Results and Analysis}
Performance of the proposed deep learning based interpolation method in video coding for classes B, C, D, E and F is shown in Table \ref{tab1}. Our method has obtained on average a $1.9\%$ BD-rate saving and up to $5.6\%$ BD-rate saving for the test sequence \emph{BQTerrace}.

We also compare our method with the CNN based half-pixel interpolation method proposed in \cite{halfpel} (called CNNIF). CNNIF only replaces the half-pixel interpolation without rate-distortion optimization and is tested on HM 16.7. For fair comparison, we also test our method on HM 16.7 and only replace the half-pixel interpolation with our GVTCNN-H without rate-distortion optimization. The BD-rate reduction of the two methods for several testing classes are shown in Table \ref{tab2}. Our method still has gain over CNNIF. And the gain will be larger after integrating the GVTCNN-Q and rate-distortion optimization.

In order to further identify the effectiveness of our group variational transformation method, we additionally train the networks for each sub-pixel position at each QP separately. Average BD-rate reduction for some testing classes are shown in Table \ref{tab3}. As can be observed, results of training the networks separately and uniformly are comparable.

\section{Conclusion}
\label{sec4}
In this paper, we propose a group variational transformation deep convolutional neural network for fractional interpolation in motion compensation of video coding. The network first uniformly extracts a shared feature map from the input integer-position sample and then a group of copied shared feature maps are transformed to samples at various sub-pixel positions. Training data generation of the proposed network is also carefully analyzed and designed. Experimental results show that our method has obtained on average a $1.9\%$ BD-rate saving on the test sequences compared with HEVC. Effectiveness of the group variational transformation method adopted in our network is also well identified by experiments.

\Section{References}
\bibliographystyle{IEEEtran}
%\bibliography{bib_DCC_2018}
\bibliography{refs}

\end{document}